\documentclass[sigconf]{acmart}

\usepackage{subcaption}
\usepackage{adjustbox}

\makeatletter
\renewcommand{\@captionheadfont}{\small}  
\renewcommand{\@captionfont}{\small}       
\makeatother

\copyrightyear{2026}
\acmYear{2026}
\setcopyright{cc}
\setcctype{by} 
\acmConference[WWW '26]{Proceedings of the ACM Web Conference 2026}{April 13--17, 2026}{Dubai, United Arab Emirates}
\acmBooktitle{Proceedings of the ACM Web Conference 2026 (WWW '26), April 13--17, 2026, Dubai, United Arab Emirates}
\acmDOI{10.1145/3774904.3792873}
\acmISBN{979-8-4007-2307-0/2026/04}
\begin{document}

\title{HierCon: Hierarchical Contrastive Attention \\for Audio Deepfake Detection}

\author{Zhili Nicholas Liang}
\orcid{0009-0005-3553-2730}
\email{nnliang@student.}
\email{unimelb.edu.au}
\affiliation{%
  \institution{University of Melbourne}
  \city{Melbourne}
  \state{Victoria}
  \country{Australia}
}

\author{Soyeon Caren Han}
\orcid{0000-0002-1948-6819}
\email{caren.han@unimelb.edu.au}
\affiliation{%
  \institution{University of Melbourne}
  \city{Melbourne}
  \state{Victoria}
  \country{Australia}
}

\author{Qizhou Wang}
\orcid{0009-0005-5097-5847}
\email{mike.wang@unimelb.edu.au}
\affiliation{%
  \institution{University of Melbourne}
  \city{Melbourne}
  \state{Victoria}
  \country{Australia}
}

\author{Christopher Leckie}
\orcid{0000-0002-4388-0517}
\email{caleckie@unimelb.edu.au}
\affiliation{%
  \institution{University of Melbourne}
  \city{Melbourne}
  \state{Victoria}
  \country{Australia}
}

\renewcommand{\shortauthors}{Liang, et al.}

\begin{abstract}
Audio deepfakes generated by modern TTS and voice conversion systems are increasingly difficult to distinguish from real speech, raising serious risks for security and online trust. While state-of-the-art self-supervised models provide rich multi-layer representations, existing detectors treat layers independently and overlook temporal and hierarchical dependencies critical for identifying synthetic artefacts. We propose HierCon, a hierarchical layer attention framework combined with margin-based contrastive learning that models dependencies across temporal frames, neighbouring layers, and layer groups, while encouraging domain-invariant embeddings. Evaluated on ASVspoof 2021 DF and In-the-Wild datasets, our method achieves state-of-the-art performance (1.93\% and 6.87\% EER), improving over independent layer weighting by 36.6\% and 22.5\% respectively. The results and attention visualisations confirm that hierarchical modelling enhances generalisation to cross-domain generation techniques and recording conditions. Source code is available at https://github.com/adlnlp/HierCon.
\end{abstract}


\settopmatter{authorsperrow=4}
\maketitle

\section{Introduction}
Recent progress in Text-to-Speech (TTS) and Voice Conversion (VC) systems has made synthetic speech increasingly realistic, enabling attackers to generate speech that closely mimics genuine human voices~\cite{muller2022does}. Such advances raise significant security and trust concerns in voice authentication, digital forensics, and online communication~\cite{jung2022aasist}. Early deepfake detection relied on handcrafted spectral or cepstral features~\cite{todisco2018integrated}, but the field has rapidly shifted toward deep learning architectures~\cite{ravanelli2018speaker,jung2022aasist} and self-supervised learning (SSL) models~\cite{chen2022wavlm,babu2022xlsr}, which offer richer multi-layer speech representations. Among these, XLS-R-based approaches have demonstrated strong results; for example, Sensitive Layer Selection (SLS) achieves 2.09\% EER on ASVspoof 2021 DF by learning scalar weights across transformer layers~\cite{zhang2024audio}.
However, methods such as SLS treat each transformer layer independently, overlooking two key sources of discriminative information. First, temporal dynamics within layers are ignored so every frame receives equal weight despite the fact that only certain regions carry synthesis artefacts. Second, inter-layer dependencies are not modelled—shallow layers encode acoustic cues while deeper layers capture semantic or prosodic signals, and their interactions are often essential for detecting modern deepfakes. Ignoring these dependencies can lead to feature homogenisation and poorer generalisation to cross-domain generation pipelines~\cite{xiao2025layer}.

To address these limitations, we introduce a hierarchical layer attention framework with margin-based contrastive learning. Our approach explicitly models dependencies at three levels: temporal attention within each layer, intra-group attention across neighbouring layers, and inter-group attention across broader layer clusters. We incorporate contrastive regularisation to encourage domain-invariant embedding geometry, enabling better generalisation to diverse deepfake generation methods. Evaluated on ASVspoof 2021 DF and In-the-Wild benchmarks, our method achieves state-of-the-art performance, reducing EER to 1.93\% and 6.87\% respectively, representing 36.6\% and 22.5\% relative improvements over independent layer weighting.
Our contributions are summarised as follows:

\noindent\textbf{1) Hierarchical Layer Attention Framework}: We propose a three-stage attention architecture that models temporal, intra-layer, and inter-layer dependencies, allowing the model to exploit complementary cues across the full representational hierarchy rather than treating layers independently.
\textbf{2) Contrastive Regularisation for Domain-Invariant Learning}: We introduce margin-based contrastive learning during fine-tuning to impose explicit geometric constraints on layer representations. Unlike prior SSL-based detectors~\cite{zhang2024audio,xiao2025layer}, our approach jointly optimises classification and contrastive objectives, improving cross-domain robustness and preventing reliance on dataset-specific artefacts.
\textbf{3) Interpretability and Robust Generalisation}: Our hierarchical attention structure naturally provides interpretable insights into which temporal regions and layer groups drive predictions. Experiments on ASVspoof 2021 DF and In-the-Wild demonstrate strong generalisation and outperform prior layer-fusion approaches.

\section{Proposed Method}
\subsection{Preliminaries}
\textbf{SSL-based Audio Deepfake Detection.}
A popular line of detection methods leverages pretrained SSL audio models and has shown strong performance. Among these, XLS-R is the most common choice. It is a large-scale cross-lingual representation model built on Wav2Vec 2.0 and trained on 128 languages to capture universal acoustic and linguistic features. To encode a raw waveform, it first extracts frame level representations at predefined intervals, which are then processed by a $L$-layer transformer network to obtain contextualised hidden states $\mathbf{H} = [\mathbf{h}_1, \mathbf{h}_2, \ldots, \mathbf{h}_L]$.

\noindent\textbf{Leveraging Multi-layer Representations.}
Existing SSL-based detectors typically utilise all hidden states to capture a more comprehensive range of audio features across multiple levels, including shallow acoustic, mid-level prosodic and deep semantic features. However, a key limitation of these methods is they assume all hidden states carry equally important information that can be summarised through uniform weighting. This can dilute discriminative signals, and ignoring inter-layer dependencies limits the ability to capture the complementary relationships across and within each level. This motivates HierCon, which dynamically models these dependencies across multiple granularities.

\subsection{HierCon}
To fully capture discriminative information from XLS-R features, we introduce \textbf{Hier}archical \textbf{Con}trastive Attention (HierCon). At its core is a hierarchical attention module that groups XLS-R's 24 layers into 8 groups of 3 consecutive layers. This grouping is motivated by the observation that neighbouring transformer layers capture similar abstraction levels~\cite{xiao2025layer}, enabling \textit{intra-group attention} to model local dependencies and \textit{inter-group attention} to integrate complementary evidence across the hierarchy (early acoustic, mid-level prosodic, high-level semantic). Combined with margin-based contrastive learning, this design learns generalisable spoofing patterns rather than dataset-specific artefacts.

\begin{figure}[t]
    \centering
    \includegraphics[width=\columnwidth]{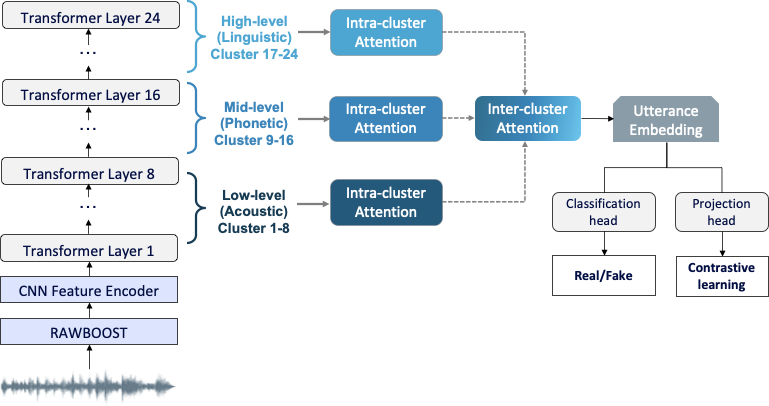}
    \caption{Overall Architecture of the Proposed HierCon, a Hierarchical Contrastive Attention for the Reliable Audio Deepfake Detection.}
    \label{fig:architecture}
\vspace{-0.5em}
\end{figure}

\vspace{1em}
\noindent\textbf{1) Hierarchical Attention.}
HierCon is designed to capture dependencies hierarchically through three stages by applying temporal attention within each layer, intra-group attention among neighbouring layers, and inter-group attention across layer groups. 

In \textit{Stage 1. Temporal Attention}, in order to emphasise informative frames capturing discriminative artefacts while down-weighting benign content, each layer $l$ applies learnable attention over $T$ temporal frames. Given frame-level hidden states $\mathbf{h}_t \in \mathbb{R}^{1024}$ for $t = 1, \ldots, T$, we compute attention weights via a two-layer MLP with learnable parameters $\mathbf{W}_1 \in \mathbb{R}^{128 \times 1024}$, $\mathbf{b}_1 \in \mathbb{R}^{128}$, and $\mathbf{w}_2 \in \mathbb{R}^{128}$:
\begin{equation}
\mathbf{e}_t = \tanh(\mathbf{W}_1 \mathbf{h}_t + \mathbf{b}_1), \quad
\alpha_t = \text{softmax}(\mathbf{w}_2^\top \mathbf{e}_t), \quad
\mathbf{z}_l = \sum_{t=1}^T \alpha_t \mathbf{h}_t
\end{equation}
yielding layer tokens $\mathbf{Z} = [\mathbf{z}_1, \ldots, \mathbf{z}_{24}] \in \mathbb{R}^{24 \times 1024}$.
In \textit{Stage 2. Intra-Group Attention}, in order to exploit complementary information within similar abstraction levels, we partition 24 layers into eight groups of three ($g=3$):
$\mathbf{z}_k' = \text{AttnPool}(\mathbf{G}_k) + \text{MLP}(\text{AttnPool}(\mathbf{G}_k))$,
where $\mathbf{G}_k \in \mathbb{R}^{3 \times 1024}$ contains neighbouring layer tokens for group $k \in \{1, \ldots, 8\}$. Early groups emphasise acoustic artefacts while late groups capture semantic cues.

In \textit{Stage 3. Inter-Group Attention}, different generation techniques manifest artefacts at different abstraction levels. Aggregating refined group vectors $\{\mathbf{z}_k'\}_{k=1}^8$ yields utterance embedding:
\begin{align}
\mathbf{u} = \text{AttnPool}(\{\mathbf{z}_k'\}) + \text{MLP}(\text{AttnPool}(\{\mathbf{z}_k'\}))
\end{align}
Adaptive weights identify which abstraction levels provide discriminative evidence for each sample.

\vspace{0.25em}
\noindent\textbf{2) Classifier and Projection Head}
The classifier is a regularised MLP (dropout + layer norm) mapping $\mathbf{u} \in \mathbb{R}^{1024}$ to logits $\mathbf{y} \in \mathbb{R}^2$. In parallel, a projection head maps $\mathbf{u}$ to a 256-dimensional space for contrastive learning, yielding embeddings $\mathbf{f}_i \in \mathbb{R}^{256}$. The lower-dimensional projection creates a geometric bottleneck encouraging compact, discriminative representations while preventing interference between classification and contrastive objectives. The projection head is discarded at inference.

\vspace{0.25em}
\noindent\textbf{3) Loss Functions}
We jointly use a binary cross-entropy (BCE) loss and a margin-based contrastive loss to stabilise training and shape the embedding geometry. The BCE term drives accurate real-vs-fake classification, while the contrastive term mitigates feature collapse from entropy-based training and reduces overfitting to dataset-specific artefacts.
Given a batch of $N$ samples with utterance embeddings $\{\mathbf{f}_i\}_{i=1}^N$, we first project them into a 256-dimensional space and compute cosine similarities. For each anchor $i$, we define the positive set $\mathcal{P}_i$ as samples from the same class (real or fake) and the negative set $\mathcal{N}_i$ as samples from the opposite class. We then compute the mean similarities
\begin{equation}
\bar{s}_i^{(+)} = \frac{1}{|\mathcal{P}_i|} \sum_{j \in \mathcal{P}_i} \cos(\mathbf{f}_i, \mathbf{f}_j), \quad
\bar{s}_i^{(-)} = \frac{1}{|\mathcal{N}_i|} \sum_{j \in \mathcal{N}_i} \cos(\mathbf{f}_i, \mathbf{f}_j),
\end{equation}
and enforce a margin $m$ between positive and negative similarities:
\begin{equation}
\mathcal{L}_{\text{con}} = \frac{1}{N} \sum_{i=1}^N \max\bigl(0,\, m + \bar{s}_i^{(-)} - \bar{s}_i^{(+)}\bigr).
\end{equation}

\noindent The overall training objective is
\begin{equation}
\mathcal{L} = \mathcal{L}_{\text{CE}} + \lambda_{\text{con}} \mathcal{L}_{\text{con}},
\end{equation}
where $\mathcal{L}_{\text{CE}}$ denotes the BCE loss and $\lambda_{\text{con}}$ controls the strength of contrastive regularisation. We set $\lambda_{\text{con}} = 0.1$ based on validation experiments: larger values (0.5–1.0) overemphasise the contrastive term and degrade classification accuracy, while smaller values ($<0.05$) yield only marginal improvements. We keep this weighting constant rather than scheduled so that the representation space is shaped jointly by both objectives from the beginning of training, avoiding convergence to suboptimal geometries.

\vspace{0.25em}
\noindent\textbf{4) Interpretability}
The hierarchical attention mechanism provides interpretability by revealing which temporal frames, layers, and groups contribute most to detection. Attention weights ($\alpha_t$, $\beta_k$, $\gamma$) can be visualised as heatmaps to identify discriminative temporal segments and layer hierarchies.

\section{Experiments}
We evaluate HierCon across three dimensions: (i) overall detection performance compared to state-of-the-art baselines, (ii) ablation analysis to isolate the contributions of hierarchical attention and contrastive learning, and (iii) interpretability analysis through multi-stage attention visualisation.  

\noindent\textbf{Experimental Setup}
All models are trained using the ASVspoof 2019 LA subset and evaluated on three benchmarks with varying levels of cross-domain difficulty: ASVspoof 2021 LA (148,176 utterances), ASVspoof 2021 DF (533,928 utterances spanning over 100 deepfake generation pipelines), and In-the-Wild (ITW) (31,779 real-world samples). Detection performance is reported using Equal Error Rate (EER\%), the standard metric in audio anti-spoofing research.
We adopt XLS-R 300M as our feature backbone, extracting hidden states from all 24 transformer layers and segmenting audio into 4-second windows. HierCon groups layers into clusters of three (8 groups total) and applies temporal, intra-group, and inter-group attention with dimensions 128 and 512 for attention and feed-forward components, respectively.
Training uses the Adam optimiser with learning rate $1 \times 10^{-6}$, batch size 16, and early stopping over 50 epochs. We apply RawBoost augmentation~\cite{tak2022rawboost} and average all reported results over three randomised seeds. Contrastive learning uses batch-level sampling, where positive pairs share spoofing class (real $\longleftrightarrow$ real or fake $\longleftrightarrow$ fake) and negative pairs differ. Average similarity scores form batch-level positive and negative constraints in the contrastive margin objective.

\subsection{Overall Performance}
\begin{figure*}[t]
\centering
\begin{subfigure}[b]{0.32\textwidth}
    \centering
\includegraphics[width=\textwidth]{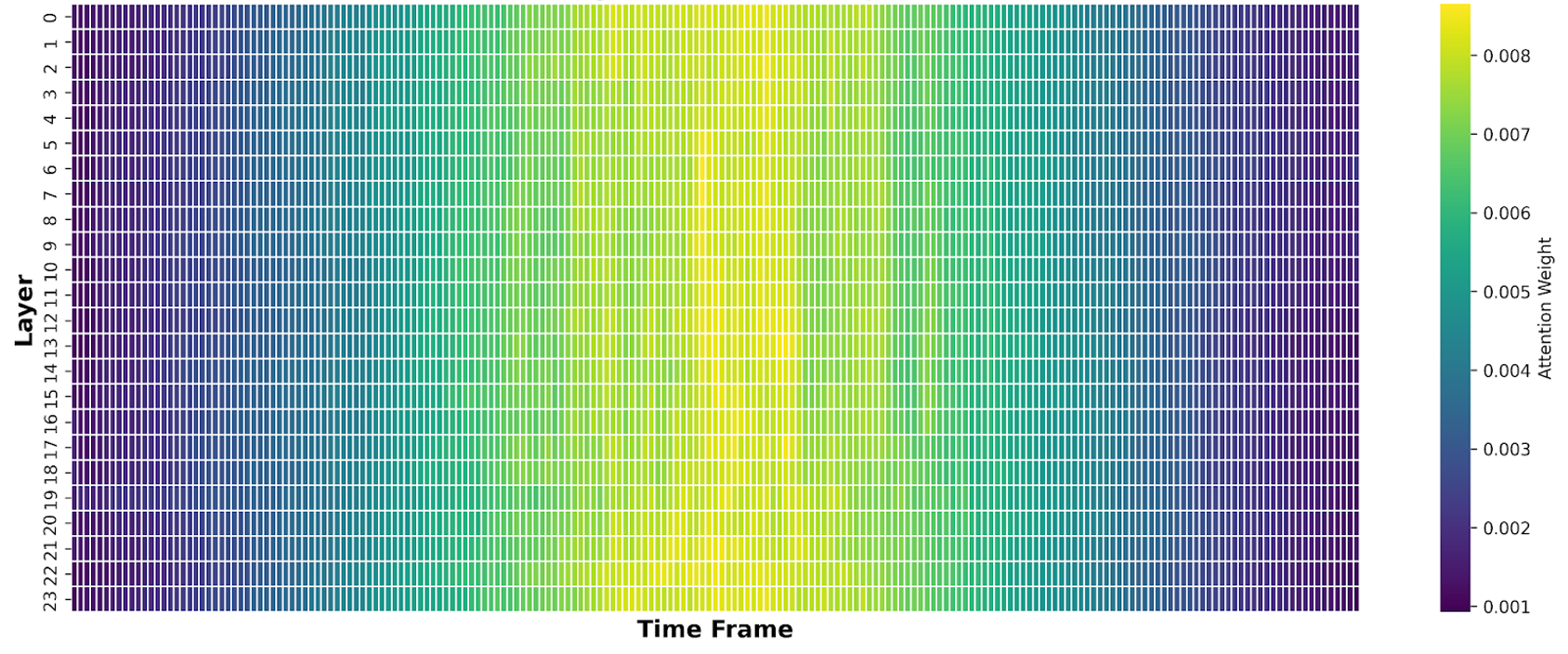}
    \caption{Temporal Attention (Stage 1)}
    \label{fig:temporal_attention}
\end{subfigure}
\hfill
\begin{subfigure}[b]{0.28\textwidth}
    \centering
    \includegraphics[width=\textwidth]{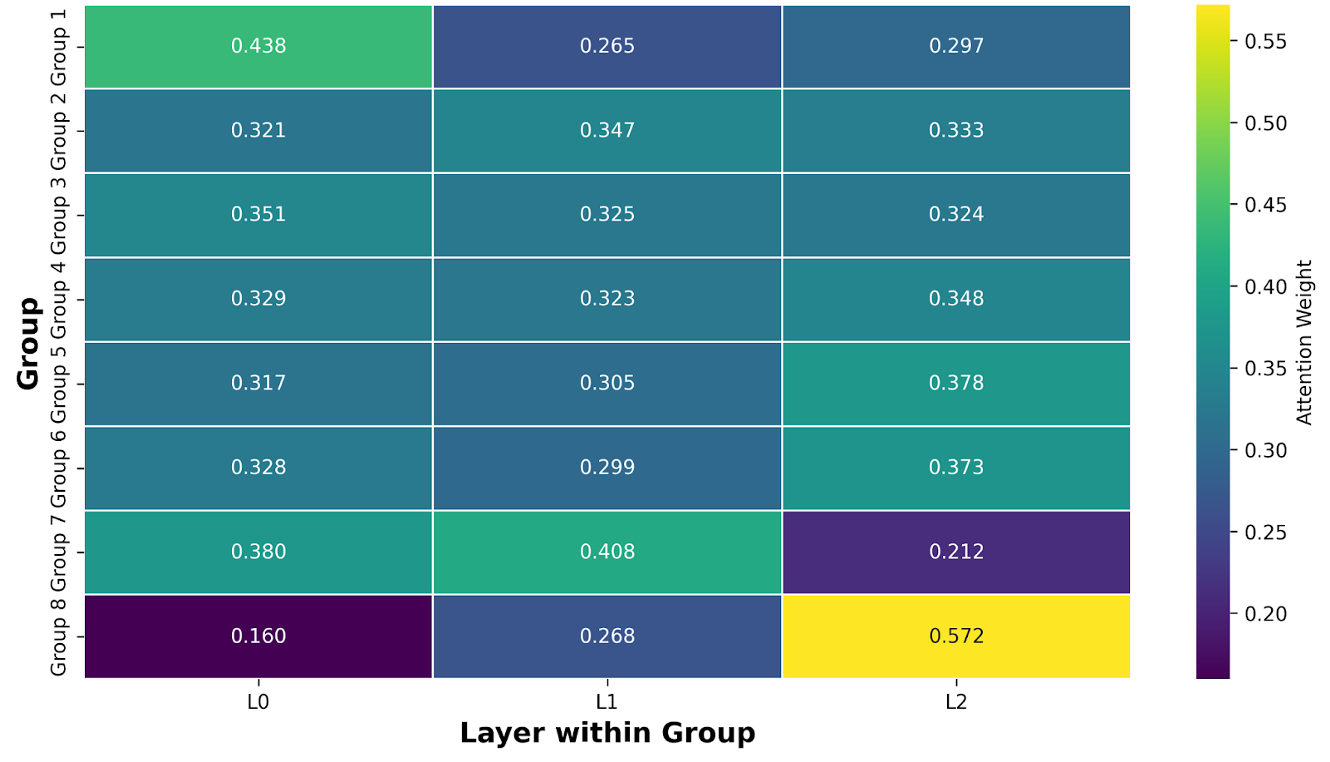}
    \caption{Intra-Group Attention (Stage 2)}
    \label{fig:intra_attention}
\end{subfigure}
\hfill
\begin{subfigure}[b]{0.32\textwidth}
    \centering
    \includegraphics[width=\textwidth]{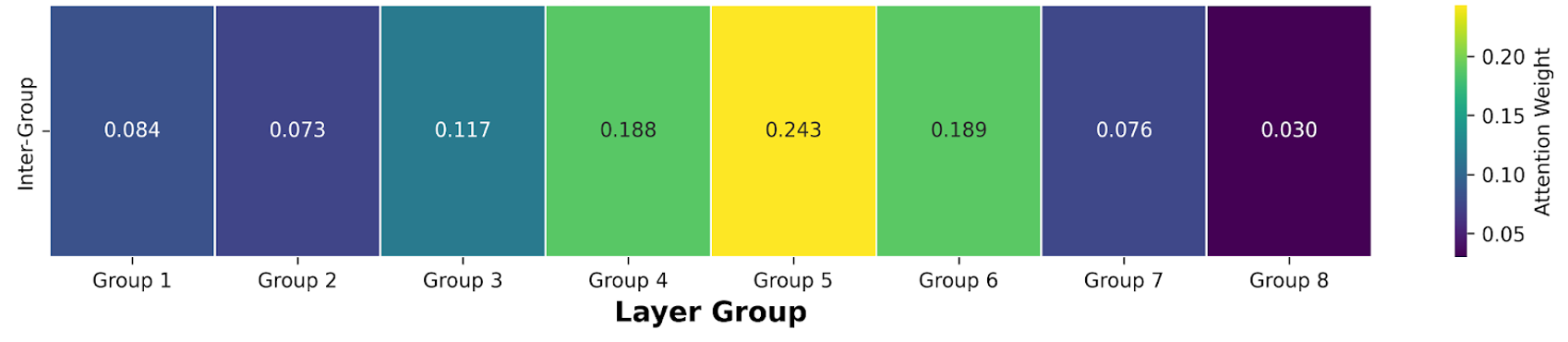}
    \caption{Inter-Group Attention (Stage 3)}
    \label{fig:inter_attention}
\end{subfigure}
\vspace{-0.5em}
\caption{HierCon attention averaged over 200 DF samples: (a) temporal focuses on mid frames (40--70\%); (b) intra-group shifts from shallow (L0: 0.438) to deep (L2: 0.572); (c) inter-group peaks at Group 5 (L12--14: 0.243).}
\vspace{-0.5em}
\label{fig:attention_composite}
\end{figure*}

Table~\ref{tab:main_results} presents pooled EERs across the three datasets with comparisons to recent detector families, including Wav2Vec/WavLM-based systems, one-class detection frameworks, and the self-supervised XLS-R SLS~\cite{zhang2024audio} baseline. 
HierCon establishes new state-of-the-art performance trends under cross-domain evaluation: 

\noindent\textbf{21 DF (Deepfake)}: HierCon achieves 1.93\% EER, surpassing the strong XLS-R + SLS baseline (2.09\%) and outperforming OCKD (2.27\%) and WavLM-based systems. The improvement is particularly impactful given the diversity of DF generation techniques.

\noindent\textbf{21 LA}: With 2.46\% EER, HierCon narrows the gap with highly specialised artefact-aware detectors (such as AASIST), while improving substantially over the XLS-R baseline (3.88\%).

\noindent\textbf{In-the-Wild}: HierCon reduces EER to 6.87\%, a 22.5 \% relative improvement over XLS-R + SLS (8.87\%).



\begin{table}[t]
\centering
\caption{Pooled EER (\%) on ASVspoof 2021 benchmarks. Results averaged over 3 runs. Bold indicates the best performance. $^\dagger$ indicates results reported from original papers.}
\vspace{-5pt}
\label{tab:main_results}
\begin{adjustbox}{width=0.95\linewidth}
\begin{tabular*}{\columnwidth}{@{\extracolsep{\fill}}lccc@{}}
\toprule
\textbf{Model} & \textbf{21 LA} & \textbf{21 DF} & \textbf{ITW} \\
\midrule
Wav2Vec+LogReg~\cite{baevski2020wav2vec}$^\dagger$ & - & - & 7.20 \\
WavLM+ASP~\cite{chen2022wavlm}$^\dagger$  & 3.31 & 4.47 & -\\
Wav2Vec+Linear~\cite{baevski2020wav2vec}$^\dagger$  & 3.63 & 3.65 & 16.17\\
WavLM+AttM~\cite{chen2022wavlm}$^\dagger$ & 3.50 & 3.19 & - \\
Wav2Vec+AASIST~\cite{jung2022aasist}$^\dagger$  & 0.82 & 2.85 & -\\
FTDKD~\cite{sahidullah2015comparison}$^\dagger$ & 2.96 & 2.82 & - \\
Wav2Vec+AASIST2~\cite{jung2022aasist}$^\dagger$  & 1.61 & 2.77 & -\\
WavLM+MFA~\cite{chen2022wavlm}$^\dagger$ & 5.08 & 2.56 & - \\
OCKD~\cite{lu2024one}$^\dagger$  & 0.90 & 2.27 & 7.86\\
OC+ACS~\cite{lavrentyeva2019stc}$^\dagger$  & 1.30 & 2.19 & -\\
XLS-R + SLS~\cite{zhang2024audio} & 3.88 & 2.09 & 8.87 \\
\midrule
\textbf{XLS-R + HierCon (Ours)} & \textbf{2.46} & \textbf{1.93} & \textbf{6.87} \\
\bottomrule
\end{tabular*}
\end{adjustbox}
\end{table}

\begin{table}[t]
\centering
\caption{Ablation study isolating contributions of hierarchical attention and contrastive learning. Bold shows the best.}
\vspace{-5pt}
\label{tab:ablation}
\begin{adjustbox}{width=0.95\linewidth}
\begin{tabular}{lccc}
\toprule
\textbf{Model} & \textbf{21 LA} & \textbf{21 DF} & \textbf{ITW} \\
\midrule
XLS-R + SLS~\cite{zhang2024audio} (Baseline) & 3.88 & 2.09 & 8.87 \\
\midrule
+ Hierarchical Attention & 2.97 & 2.13 & 8.81 \\
+ Hier. Attn + Contrastive Learning & \textbf{2.46} & \textbf{1.93} & \textbf{6.87} \\
\bottomrule
\end{tabular}
\end{adjustbox}
\vspace{-1em}
\end{table}

\subsection{Ablation Studies}
To quantify component-level contributions, we incrementally activate HierCon modules beginning from the XLS-R + SLS baseline. Results are summarised in Table ~\ref{tab:ablation}. 
The baseline performs reasonably well in-domain but struggles under distribution shift, particularly in ITW, indicating its limited ability to capture robust generative artefacts beyond fixed layer-level weighting.
Adding hierarchical attention without contrastive learning provides substantial gains on cross-domain datasets: 23.5\% relative improvement on LA (2.97\% vs 3.88\%) and a modest 0.7\% improvement on In-the-Wild (8.81\% vs 8.87\%). However, performance on DF slightly degrades (2.13\% vs 2.09\%), suggesting hierarchical modelling effectively captures inter-layer dependencies for generalisation but may overfit to dataset-specific statistical cues when not regularised. This behaviour indicates that while structural modelling strengthens feature abstraction, it requires additional stabilisation to remain resilient across unseen attack paradigms.
The combination of hierarchical attention with margin-based contrastive learning yields the best performance across all datasets, with dramatic improvements on the most challenging generalisation benchmark: 27.7\% relative gain over SLS on In-the-Wild (6.87\% vs 8.87\%). Notably, contrastive learning also alleviates the DF degradation (1.93\% vs 2.13\%), demonstrating its role in enforcing a more transferable representation structure. The gains imply that contrastive regularisation not only strengthens class separation but also prevents the hierarchical mechanism from relying on shallow artefacts inconsistent across spoofing algorithms or recording environments.
These findings validate the dual-objective design. Hierarchical attention captures complementary cues distributed across transformer layers, while contrastive learning encourages more discriminative and less dataset-dependent representations. The consistent improvements suggest architectural and training components jointly contribute to stronger generalisation, enabling reliable deepfake detection.

\vspace{-0.5em}
\subsection{Attention Analysis and Interpretability}
To evaluate whether the learned behavior aligns with the intended design, we analyse attention distributions across the three HierCon stages. As can be seen in Figure~\ref{fig:attention_composite}, temporal attention consistently prioritises central regions of the signal while down-weighting boundaries, a pattern consistent with established forensic observations that synthetic artefacts are more likely to appear during sustained speech rather than at the onset or trailing edges. The consistency of this behavior across runs suggests that the model is learning stable temporal cues rather than relying on optimisation noise or sample-specific alignment bias.
At the intra-group stage, we observe a gradual shift from emphasizing shallow, acoustically driven layers toward deeper, semantically oriented layers as processing transitions across groups. This progression reflects how modern generative systems introduce artefacts at multiple abstraction levels: shallow layers tend to expose subtle spectral distortions or noise shaping irregularities, while deeper layers capture unnatural prosody, rhythm, or linguistic instabilities introduced by synthesis pipelines. Meanwhile, inter-group attention places the highest emphasis on mid-level transformer layers. This suggests that the most discriminative evidence emerges from the interaction zone between purely acoustic and purely semantic representations, aligning with prior findings in multi-layer self-supervised feature fusion research.
Beyond structural inspection, we also observe that attention patterns remain highly consistent across different spoofing mechanisms, including vocoder-based, diffusion-based, and voice conversion models. This convergence implies that the model is not simply memorizing synthesis-specific signatures but instead identifying recurrent artefact structures that persist despite variation in generation pipelines. The stability of these patterns across domains supports the hypothesis that HierCon learns transferable cues grounded in the fundamental properties of synthetic speech.
Taken together, the interpretability analysis indicates that HierCon improves performance not through opaque shortcut features but through structured, coherent use of multi-scale representations aligned with known spoofing signal characteristicsThe observed behaviours demonstrate that the model develops reasoning across abstraction levels rather than depending solely on localised artefacts, reinforcing confidence in its resilience and generalisation beyond controlled benchmark datasets.

\section{Conclusion}
We introduced HierCon, a hierarchical attention framework with contrastive learning for audio deepfake detection. The method addresses limitations of prior SSL-based approaches by modeling inter-layer dependencies and enforcing domain-invariant embedding structure.
Across benchmarks, HierCon achieves state-of-the-art performance, including 1.93\% EER on DF and a 22.5\% relative improvement over XLS-R+SLS~\cite{zhang2024audio}. Ablations confirm that hierarchical attention and contrastive learning contribute complementary benefits, enabling both improved generalisation and transparent decision behavior.
Future work will examine cross-modal and deployment-efficient variants suitable for real-time forensic use.

\section{Acknowledgments}
This research was supported by the Korea Planning \& Evaluation Institute of Industrial Technology (KEIT) funded by the Ministry of Trade, Industry and Energy (No.RS-2025-25458052, Development of Core Technologies for Manufacturing Foundation Models) and the Institute of Information \& communications Technology Planning \& Evaluation (IITP) grant funded by the Korea government(MSIT) (No.RS-2025-02217259, Development of self-evolving AI bias detection-correction-explain platform based on international multidisciplinary governance).

\smallskip

\noindent Qizhou Wang and Christopher Leckie were supported in part by the Australian Internet Observatory (AIO), a national research infrastructure supporting digital platform and smart data research. AIO received investment from the Australian Research Data Commons (ARDC) through the National Collaborative Research Infrastructure Strategy (NCRIS).

\vspace{0.8cm}

\bibliographystyle{ACM-Reference-Format}
\balance
\bibliography{ref}


\begin{thebibliography}{13}


\ifx \showCODEN    \undefined \def \showCODEN     #1{\unskip}     \fi
\ifx \showISBNx    \undefined \def \showISBNx     #1{\unskip}     \fi
\ifx \showISBNxiii \undefined \def \showISBNxiii  #1{\unskip}     \fi
\ifx \showISSN     \undefined \def \showISSN      #1{\unskip}     \fi
\ifx \showLCCN     \undefined \def \showLCCN      #1{\unskip}     \fi
\ifx \shownote     \undefined \def \shownote      #1{#1}          \fi
\ifx \showarticletitle \undefined \def \showarticletitle #1{#1}   \fi
\ifx \showURL      \undefined \def \showURL       {\relax}        \fi
\providecommand\bibfield[2]{#2}
\providecommand\bibinfo[2]{#2}
\providecommand\natexlab[1]{#1}
\providecommand\showeprint[2][]{arXiv:#2}

\bibitem[Babu et~al\mbox{.}(2022)]%
        {babu2022xlsr}
\bibfield{author}{\bibinfo{person}{A. Babu}, \bibinfo{person}{C. Wang}, \bibinfo{person}{A. Tjandra}, \bibinfo{person}{K. Lakhotia}, \bibinfo{person}{Q. Xu}, \bibinfo{person}{N. Goyal}, \bibinfo{person}{K. Singh}, \bibinfo{person}{P. von Platen}, \bibinfo{person}{Y. Saraf}, \bibinfo{person}{J. Pino}, \bibinfo{person}{A. Baevski}, \bibinfo{person}{A. Conneau}, {and} \bibinfo{person}{M. Auli}.} \bibinfo{year}{2022}\natexlab{}.
\newblock \showarticletitle{{XLS-R}: Self-supervised cross-lingual speech representation learning at scale}. In \bibinfo{booktitle}{\emph{Interspeech}}.
\newblock


\bibitem[Baevski et~al\mbox{.}(2020)]%
        {baevski2020wav2vec}
\bibfield{author}{\bibinfo{person}{A. Baevski}, \bibinfo{person}{Y. Zhou}, \bibinfo{person}{A. Mohamed}, {and} \bibinfo{person}{M. Auli}.} \bibinfo{year}{2020}\natexlab{}.
\newblock \showarticletitle{wav2vec 2.0: A framework for self-supervised learning of speech representations}. In \bibinfo{booktitle}{\emph{NeurIPS}}. \bibinfo{pages}{12449--12460}.
\newblock


\bibitem[Chen et~al\mbox{.}(2022)]%
        {chen2022wavlm}
\bibfield{author}{\bibinfo{person}{S. Chen}, \bibinfo{person}{C. Wang}, \bibinfo{person}{Z. Chen}, \bibinfo{person}{Y. Wu}, \bibinfo{person}{S. Liu}, \bibinfo{person}{Z. Chen}, \bibinfo{person}{J. Li}, \bibinfo{person}{N. Kanda}, \bibinfo{person}{T. Yoshioka}, \bibinfo{person}{X. Xiao}, \bibinfo{person}{J. Wu}, \bibinfo{person}{L. Zhou}, \bibinfo{person}{S. Ren}, \bibinfo{person}{Y. Qian}, \bibinfo{person}{Y. Qian}, \bibinfo{person}{J. Wu}, \bibinfo{person}{M. Zeng}, \bibinfo{person}{X. Yu}, {and} \bibinfo{person}{F. Wei}.} \bibinfo{year}{2022}\natexlab{}.
\newblock \showarticletitle{{WavLM}: Large-scale self-supervised pre-training for full stack speech processing}.
\newblock \bibinfo{journal}{\emph{IEEE Journal of Selected Topics in Signal Processing}} \bibinfo{volume}{16}, \bibinfo{number}{6} (\bibinfo{year}{2022}).
\newblock


\bibitem[Jung et~al\mbox{.}(2022)]%
        {jung2022aasist}
\bibfield{author}{\bibinfo{person}{J. Jung}, \bibinfo{person}{H.-S. Heo}, \bibinfo{person}{H. Tak}, \bibinfo{person}{H. Shim}, \bibinfo{person}{J.~S. Chung}, \bibinfo{person}{B.-J. Lee}, \bibinfo{person}{H.-J. Yu}, {and} \bibinfo{person}{N. Evans}.} \bibinfo{year}{2022}\natexlab{}.
\newblock \showarticletitle{{AASIST}: Audio anti-spoofing using integrated spectro-temporal graph attention networks}. In \bibinfo{booktitle}{\emph{ICASSP}}. \bibinfo{pages}{6367--6371}.
\newblock


\bibitem[Lavrentyeva et~al\mbox{.}(2019)]%
        {lavrentyeva2019stc}
\bibfield{author}{\bibinfo{person}{G. Lavrentyeva}, \bibinfo{person}{S. Novoselov}, \bibinfo{person}{A. Malykh}, \bibinfo{person}{A. Kozlov}, \bibinfo{person}{O. Kudashev}, {and} \bibinfo{person}{V. Shchemelinin}.} \bibinfo{year}{2019}\natexlab{}.
\newblock \showarticletitle{{STC} antispoofing systems for the {ASVspoof2019} challenge}. In \bibinfo{booktitle}{\emph{Interspeech}}. \bibinfo{pages}{1033--1037}.
\newblock


\bibitem[Lu et~al\mbox{.}(2024)]%
        {lu2024one}
\bibfield{author}{\bibinfo{person}{J. Lu}, \bibinfo{person}{Y. Zhang}, \bibinfo{person}{W. Wang}, \bibinfo{person}{Z. Shang}, {and} \bibinfo{person}{P. Zhang}.} \bibinfo{year}{2024}\natexlab{}.
\newblock \showarticletitle{One-class knowledge distillation for spoofing speech detection}. In \bibinfo{booktitle}{\emph{ICASSP}}. \bibinfo{pages}{11251--11255}.
\newblock


\bibitem[M{\"u}ller et~al\mbox{.}(2022)]%
        {muller2022does}
\bibfield{author}{\bibinfo{person}{N.~M. M{\"u}ller}, \bibinfo{person}{P. Czempin}, \bibinfo{person}{F. Dieckmann}, \bibinfo{person}{A. Froghyar}, {and} \bibinfo{person}{K. B{\"o}ttinger}.} \bibinfo{year}{2022}\natexlab{}.
\newblock \showarticletitle{Does audio deepfake detection generalize?}
\newblock \bibinfo{journal}{\emph{arXiv preprint arXiv:2203.16263}} (\bibinfo{year}{2022}).
\newblock


\bibitem[Ravanelli and Bengio(2018)]%
        {ravanelli2018speaker}
\bibfield{author}{\bibinfo{person}{M. Ravanelli} {and} \bibinfo{person}{Y. Bengio}.} \bibinfo{year}{2018}\natexlab{}.
\newblock \showarticletitle{Speaker recognition from raw waveform with {SincNet}}. In \bibinfo{booktitle}{\emph{SLT}}. \bibinfo{pages}{1021--1028}.
\newblock


\bibitem[Sahidullah et~al\mbox{.}(2015)]%
        {sahidullah2015comparison}
\bibfield{author}{\bibinfo{person}{M. Sahidullah}, \bibinfo{person}{T. Kinnunen}, {and} \bibinfo{person}{C. Hanilçi}.} \bibinfo{year}{2015}\natexlab{}.
\newblock \showarticletitle{A comparison of features for synthetic speech detection}. In \bibinfo{booktitle}{\emph{Interspeech}}. \bibinfo{pages}{2087--2091}.
\newblock


\bibitem[Tak et~al\mbox{.}(2022)]%
        {tak2022rawboost}
\bibfield{author}{\bibinfo{person}{H. Tak}, \bibinfo{person}{J. Patino}, \bibinfo{person}{M. Todisco}, \bibinfo{person}{A. Nautsch}, \bibinfo{person}{N. Evans}, {and} \bibinfo{person}{A. Larcher}.} \bibinfo{year}{2022}\natexlab{}.
\newblock \showarticletitle{End-to-end anti-spoofing with {RawNet2}}. In \bibinfo{booktitle}{\emph{ICASSP}}. \bibinfo{pages}{6369--6373}.
\newblock


\bibitem[Todisco et~al\mbox{.}(2018)]%
        {todisco2018integrated}
\bibfield{author}{\bibinfo{person}{M. Todisco}, \bibinfo{person}{H. Delgado}, \bibinfo{person}{K.~A. Lee}, \bibinfo{person}{M. Sahidullah}, \bibinfo{person}{N. Evans}, \bibinfo{person}{T. Kinnunen}, {and} \bibinfo{person}{J. Yamagishi}.} \bibinfo{year}{2018}\natexlab{}.
\newblock \showarticletitle{Integrated presentation attack detection and automatic speaker verification: Common features and {Gaussian} back-end fusion}. In \bibinfo{booktitle}{\emph{Interspeech}}.
\newblock


\bibitem[Xiao and Vu(2025)]%
        {xiao2025layer}
\bibfield{author}{\bibinfo{person}{Y. Xiao} {and} \bibinfo{person}{N.~T. Vu}.} \bibinfo{year}{2025}\natexlab{}.
\newblock \showarticletitle{Layer-wise decision fusion for fake audio detection using {XLS-R}}. In \bibinfo{booktitle}{\emph{Interspeech}}. \bibinfo{pages}{5618--5622}.
\newblock


\bibitem[Zhang et~al\mbox{.}(2024)]%
        {zhang2024audio}
\bibfield{author}{\bibinfo{person}{Q. Zhang}, \bibinfo{person}{S. Wen}, {and} \bibinfo{person}{T. Hu}.} \bibinfo{year}{2024}\natexlab{}.
\newblock \showarticletitle{Audio deepfake detection with self-supervised {XLS-R} and {SLS} classifier}. In \bibinfo{booktitle}{\emph{ACM MM}}. \bibinfo{pages}{6765--6773}.
\newblock


\end{thebibliography}

\end{document}